\documentclass[slac_one]{revtex4}
\usepackage{graphicx}
\usepackage{epsfig}
\usepackage{fancyhdr}
\usepackage{amsfonts}
\usepackage{amsmath}
\usepackage{amssymb}
\usepackage{axodraw}
\usepackage[dvips]{color}
\usepackage{colordvi,psfrag}
\makeatletter
\def\citer{\@ifnextchar [{\@tempswatrue\@citexr}{\@tempswafalse\@citexr[]}}
 
\def\@citexr[#1]#2{\if@filesw\immediate\write\@auxout{\string\citation{#2}}\fi
  \def\@citea{}\@cite{\@for\@citeb:=#2\do
    {\@citea\def\@citea{--\penalty\@m}\@ifundefined
       {b@\@citeb}{{\bf ?}\@warning
       {Citation `\@citeb' on page \thepage \space undefined}}%
\hbox{\csname b@\@citeb\endcsname}}}{#1}}
\makeatother

\def\refeq#1{\mbox{Eq.~(\ref{#1})}}

\def\reffi#1{\mbox{Fig.~\ref{#1}}}

\def\refta#1{\mbox{Tab.~\ref{#1}}}
\def\citere#1{\mbox{Ref.~\cite{#1}}}
\def\citeres#1{\mbox{Refs.~\cite{#1}}}

\def\xtilde#1{%
  \setbox0\hbox{$\tilde#1$}%
  \rlap{\raise\ht0\hbox{\tiny$_{\,(\;\,)}$}}%
  \tilde#1%
}

\newcommand{\mste}{m_{\tilde{t}_1}}
\newcommand{\mstz}{m_{\tilde{t}_2}}

\newcommand{\msbar}{$\overline{\rm{MS}}$}

\def\order#1{${\cal O}(#1)$}
\newcommand{\cp}{{\cal CP}}

\newcommand{\MW}{M_W}
\newcommand{\MZ}{M_Z}
\newcommand{\MA}{M_A}
\newcommand{\mh}{m_h}

\newcommand{\mhmax}{m_h^{\rm max}}

\newcommand{\mt}{m_{t}}
\newcommand{\mtexp}{m_{t}^{\rm exp}}

\newcommand{\tsf}{\theta\kern-.20em_{\tilde{f}}}
\newcommand{\tsfp}{\theta\kern-.20em_{\tilde{f}\prime}}
\newcommand{\tsq}{\theta\kern-.15em_{\tilde{q}}}

\newcommand{\sweff}{\sin^2\theta_{\mathrm{eff}}}

\newcommand{\KL}{\left(}
\newcommand{\KR}{\right)}

\newcommand{\VL}{\left( \begin{array}{c}}
\newcommand{\VR}{\end{array} \right)}
\newcommand{\ML}{\left( \begin{array}{cc}}
\newcommand{\MLd}{\left( \begin{array}{ccc}}
\newcommand{\MLv}{\left( \begin{array}{cccc}}
\newcommand{\MR}{\end{array} \right)}

\newcommand{\tb}{\tan \beta}

\newcommand{\SQb}{\sin^2\beta\hspace{1mm}}

\newcommand{\gev}{\,\, {\rm GeV}}
\newcommand{\mev}{\,\, {\rm MeV}}

\newcommand{\BC}{\begin{center}}
\newcommand{\EC}{\end{center}}
\newcommand{\BE}{\begin{equation}}
\newcommand{\EE}{\end{equation}}
\newcommand{\BEA}{\begin{eqnarray}}
\newcommand{\BEAnn}{\begin{eqnarray*}}
\newcommand{\EEA}{\end{eqnarray}}
\newcommand{\EEAnn}{\end{eqnarray*}}

\newcommand{\id}{{\rm 1\kern-.12em
\rule{0.3pt}{1.5ex}\raisebox{0.0ex}{\rule{0.1em}{0.3pt}}}}
\newcommand{\lsim}
{\;\raisebox{-.3em}{$\stackrel{\displaystyle <}{\sim}$}\;}

\newcommand{\gf}{G_F}

\def\al{\alpha}

\def\de{\delta}

\def\De{\Delta}

\newcommand{\plane}[2]{$#1$--$#2$~plane}

\def\3{\ss}

\begin{document}

\title{{\small{2005 International Linear Collider Workshop - Stanford,
U.S.A.}}\\ %% Please keep this conference title here
\vspace{12pt}
Physics Gain of a Precise \boldmath{$\mt$} Measurement} %% 


%




\author{S.~Heinemeyer}
\affiliation{CERN TH Division, Department of Physics,
CH-1211 Geneva 23, Switzerland}
\author{G.~Weiglein}
\affiliation{Institute for Particle Physics Phenomenology, University
  of Durham, Durham DH1~3LE, UK} 

\begin{abstract}
The top quark mass is currently measured to 
$\de\mt^{\rm exp,Tevatron} = 2.9 \gev$ and will be measured at the LHC
to a precision of  
$\de\mt^{\rm exp,LHC} \approx \mbox{1--2} \gev$. We show that even
this impressive precision will not be sufficient for many future physics
applications. These include electroweak precision observables,
Higgs physics in extensions of the Standard Model as well as cold dark
matter predictions in Supersymmetry. The desired
experimental precision can only be reached at the ILC with 
$\de\mt^{\rm exp,ILC} \approx 100 \mev$.
\end{abstract}


\maketitle

\thispagestyle{fancy}





%%%%%%%%%%%%%%%%%%%%%%%%%%%%%%%%%%%%%%%%%%%%%%%%%%%%%%%%%%%%%%%%%%%%%%%%%%%%%%%
%%%%%%%%%%%%%%%%%%%%%%%%%%%%%%%%%%%%%%%%%%%%%%%%%%%%%%%%%%%%%%%%%%%%%%%%%%%%%%%

% Section title should be in all capitals.

%\vspace{-2cm}
\section{INTRODUCTION} 

The mass of the top quark, $\mt$, is a fundamental parameter of the 
electroweak theory. It is by far the heaviest of all quark masses and
it is also larger than the masses of all other known fundamental particles. 
The large value of $\mt$ gives rise to a large coupling between the top 
quark and the Higgs boson and is furthermore important for flavor
physics. It could therefore provide a window to new physics. The correct
prediction of $\mt$ will be a crucial test for any fundamental theory.
The top-quark mass also plays an important role in electroweak precision
physics, as a consequence in particular of non-decoupling effects being
proportional to powers of $\mt$. A precise knowledge of $\mt$ is
therefore indispensable in order to have sensitivity to possible effects
of new physics in electroweak precision tests.

The current world average for the top-quark mass from the measurement
at the Tevatron is 
$\mt = 172.7 \pm 2.9 \gev$~\cite{newmt}. 
The prospective accuracy at the LHC is 
$\de\mtexp = \mbox{1--2} \gev$~\cite{mtdetLHC}, 
while at the ILC a very precise determination of $\mt$ with an accuracy
of $\de\mtexp \lsim 100 \mev$ will be possible~\cite{lctdrs,mtdet}.
This error contains both the experimental error of the mass parameter
extracted from the $t \bar t$ threshold measurements at the ILC and 
the envisaged theoretical uncertainty from its transition into a suitable
short-distance mass (like the \msbar\ mass).

In the following we show for some examples that in many physics
applications the experimental error on $\mt$ achievable at the LHC
would be the limiting factor, demonstrating the need for the ILC
precision. More examples can be found in \citere{deltamt}.

%%%%%%%%%%%%%%%%%%%%%%%%%%%%%%%%%%%%%%%%%%%%%%%%%%%%%%%%%%%%%%%%%%%%%%%%%%%%%%%
%%%%%%%%%%%%%%%%%%%%%%%%%%%%%%%%%%%%%%%%%%%%%%%%%%%%%%%%%%%%%%%%%%%%%%%%%%%%%%%

\section{ELECTROWEAK PRECISION OBSERVABLES}

Electroweak precision observables (EWPO) can be used to
perform internal consistency checks of the model under consideration and
to obtain indirect constraints on unknown model parameters.
This is done by comparing experimental results of the EWPO with
their theory prediction within, for example, the Standard Model (SM) or
the Minimal Supersymmetric Standard Model (MSSM).

The two most prominent EWPO are the mass of the $W$~boson, $\MW$, and
the effective leptonic weak mixing angle, $\sweff$. Their current
experimental uncertainties and the prospective precisions with further
data from the Tevatron, the LHC and the ILC (including the GigaZ
option) are summarized in \refta{tab:ewpoexp},
see~\citeres{blueband,PomssmRep} for further details.

In addition to the experimental uncertainties there are two sources of
theoretical uncertainties: those from 
unknown higher-order corrections (``intrinsic'' theoretical
uncertainties), and those from experimental errors of the input
parameters (``parametric'' theoretical uncertainties). 
The current and estimated future intrinsic uncertainties within the SM
are~\cite{blueband,mwsweff}
\BEA
\De\MW^{\rm intr,today,SM} \approx 4 \mev, & \quad &
\De\sweff^{\rm intr,today,SM} \approx 5 \times 10^{-5} ~, \\
\De\MW^{\rm intr,future,SM} \approx 2 \mev, & \quad &
\De\sweff^{\rm intr,future,SM} \approx 2 \times 10^{-5} ~,
\label{eq:intruncSM}
\EEA
while in the MSSM the current intrinsic uncertainties are estimated
to~\cite{PomssmRep,drMSSMal2A,drMSSMal2B}
\BE
\De\MW^{\rm intr,today,MSSM} \approx (5 - 9) \mev, \quad
\De\sweff^{\rm intr,today,MSSM} \approx (5 - 7) \times 10^{-5} ~,
\label{eq:intruncMSSM}
\EE
depending on the supersymmetric (SUSY) mass scale. 
In the future one expects that they
can be brought down to the level of the SM, see \refeq{eq:intruncSM}. 

The parametric errors of $\MW$ and $\sweff$ induced by the top quark
mass, the uncertainty of $\De\al_{\rm had}$ (we assume a future
determination of $\de(\De\al_{\rm had}) = 5 \times 10^{-5}$~\cite{fredl})
and the experimental uncertainty of the $Z$~boson mass, 
$\de\MZ = 2.1 \mev$, are collected in \refta{tab:ewpopara}.

In order to keep the theoretical uncertainty induced by $\mt$ at a
level comparable to or smaller than the other parametric and intrinsic
uncertainties, $\de\mt$ has to be of \order{0.1 \gev} in the case of
$\MW$, and about $0.5 \gev$ in the case of $\sweff$. If the
experimental error of $\mt$ remains substantially larger, it
would constitute the limiting factor of the theoretical uncertainty.
Using the EWPO to distinguish different models from each other or to
determine indirectly the unknown model parameters the ILC precision on
$\mt$ is crucial, in particular in view of the precision measurement
of $\sweff$ at GigaZ~\cite{deltamt}. 

%%%%%%%%%%%%% T A B L E %%%%%%%%%%%%%%%%%%%%%%%%%%%%%%%%%%%%%%
\begin{table}[htb!]
\renewcommand{\arraystretch}{1.5}
\begin{center}
\caption{Expected experimental accuracies of $\MW$ and $\sweff$ at the
  Tevatron, the LHC and the ILC/GigaZ.}
\label{tab:ewpoexp}
%\hspace{-1em}
\begin{tabular}{|c||c|c|c|c|}
\cline{2-5} \multicolumn{1}{c||}{}
& ~today~ & ~Tev./LHC~ & ~ILC~  & ~GigaZ~ \\
\hline\hline
~$\de\sweff$ $[10^{-5}$]~ & 17 & 17   & --  & 1.3  \\
\hline
~$\de\MW$ [MeV]~           & 34 & 15   & 10   & 7   \\
\hline\hline
\end{tabular}
\end{center}
\end{table}
\renewcommand{\arraystretch}{1}
%%%%%%%%%%%%% T A B L E %%%%%%%%%%%%%%%%%%%%%%%%%%%%%%%%%%%%%%

%%%%%%%%%%%%% T A B L E %%%%%%%%%%%%%%%%%%%%%%%%%%%%%%%%%%%%%%
\renewcommand{\arraystretch}{1.5}
\begin{table}[htb!]
\begin{center}
\caption{Parametric errors on the prediction of $\MW$ and $\sweff$.}
\label{tab:ewpopara}
%\hspace{-1em}
\begin{tabular}{|c||c|c|c|c|c|c|}
\cline{2-7} \multicolumn{1}{c||}{}
& ~$\de\mt = 2.9 \gev$~ & ~$\de\mt = 2 \gev$~ & ~$\de\mt = 1 \gev$~ 
& ~$\de\mt = 0.1 \gev$~ & 
~$\de(\De\al_{\rm had})$~ & ~$\de\MZ$~ \\
\hline\hline
~$\de\sweff$ [$10^{-5}$]~ & $8.7$ &
$6$ & $3$ & $0.3$ & $1.8$ & $1.4$ \\
\hline
$\De\MW$ [MeV] & $17.4$ &
$12$ & $6$ & $1$ & $1$ & $2.5$ \\
\hline\hline
\end{tabular}
\end{center}
\end{table}
\renewcommand{\arraystretch}{1}
%%%%%%%%%%%%% T A B L E %%%%%%%%%%%%%%%%%%%%%%%%%%%%%%%%%%%%%%

%%%%%%%%%%%%%%%%%%%%%%%%%%%%%%%%%%%%%%%%%%%%%%%%%%%%%%%%%%%%%%%%%%%%%%%%%%%%%%%
%%%%%%%%%%%%%%%%%%%%%%%%%%%%%%%%%%%%%%%%%%%%%%%%%%%%%%%%%%%%%%%%%%%%%%%%%%%%%%%

\section{HIGGS PHYSICS IN THE MSSM AND OTHER EXTENSIONS OF THE SM}

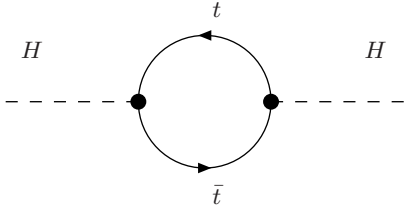
\begin{figure}[htb]
\BC
%\resizebox{13em}{!}{
\setlength{\unitlength}{1pt}
\begin{picture}(170, 70)
\DashLine(10,40)(60,40){5}
\Text(20,60)[]{$H$}
\Vertex(60,40){3}
\ArrowArc(85,40)(25,0,180)
\ArrowArc(85,40)(25,180,360)
\Text(90,75)[]{$t$}
\Text(90,5)[]{$\bar t$}
\Vertex(110,40){3}
\DashLine(110,40)(160,40){5}
\Text(150,60)[]{$H$}
\end{picture}
%}
\EC
\vspace{-2em}
\caption{Loop contribution of the top quark to the Higgs boson mass.}
\label{fig:mhiggs}
\end{figure}

Because of its large mass, the top quark is expected to have a large
Yukawa coupling to Higgs bosons, being proportional to $\mt$.
In each model where the Higgs boson mass is not a free
parameter but predicted in terms of the the other model parameters
(as e.g.\ in the MSSM), the diagram in \reffi{fig:mhiggs} contributes
to the Higgs mass. This diagram gives rise to a leading $\mt$
contribution of the form
\BE
\De\mh^2 \sim \gf \; N_C \; C \; \mt^4~,
\EE
where $\gf$ is the Fermi constant, $N_C$ is the color factor and the
coefficient $C$ depends on the specific model. Thus the experimental
error of $\mt$ necessarily leads to a parametric error in the Higgs
boson mass evaluation. 

%%%%%%%%%%%%%%%%%%%%%%%% F I G U R E %%%%%%%%%%%%%%%%%%%%%%%%%%%%%%%%%%%%%%%%%
%%%%%%%%%%%%%%%%%%%%%%%% F I G U R E %%%%%%%%%%%%%%%%%%%%%%%%%%%%%%%%%%%%%%%%%

Taking the MSSM as a specific example 
(including also the scalar top contributions and the appropriate
renormalization) $N_C \, C$ is given for the light $\cp$-even Higgs
boson mass by  
\BE
N_C \, C = \frac{3}{\sqrt{2}\,\pi^2\,\SQb} \; 
\log \KL \frac{\mste\mstz}{\mt^2} \KR~.
\EE
Here $m_{\tilde t_{1,2}}$ denote the two masses of the scalar tops.
The optimistic LHC
precision of $\de\mt = 1 \gev$ leads to an uncertainty of 
$\sim 2.5\%$ in the prediction of $\mh$, while the ILC will yield a
precision of $\sim 0.2\%$.  
These uncertainties have to be compared with the anticipated precision of
the future Higgs boson mass measurements. With a precision of 
$\de\mh^{\rm exp,LHC} \approx 0.2 \gev$~\cite{LHCHiggs} the relative
precision is at the level of $\sim 0.2\%$. It is apparent that only the
ILC precision of $\mt$ will yield a parametric error small enough to
allow a precise comparison of the Higgs boson mass prediction (where
also the intrinsic theoretical uncertainty has to be improved
accordingly) and its experimental value.

\begin{figure}[htb!]
\begin{center}
\epsfig{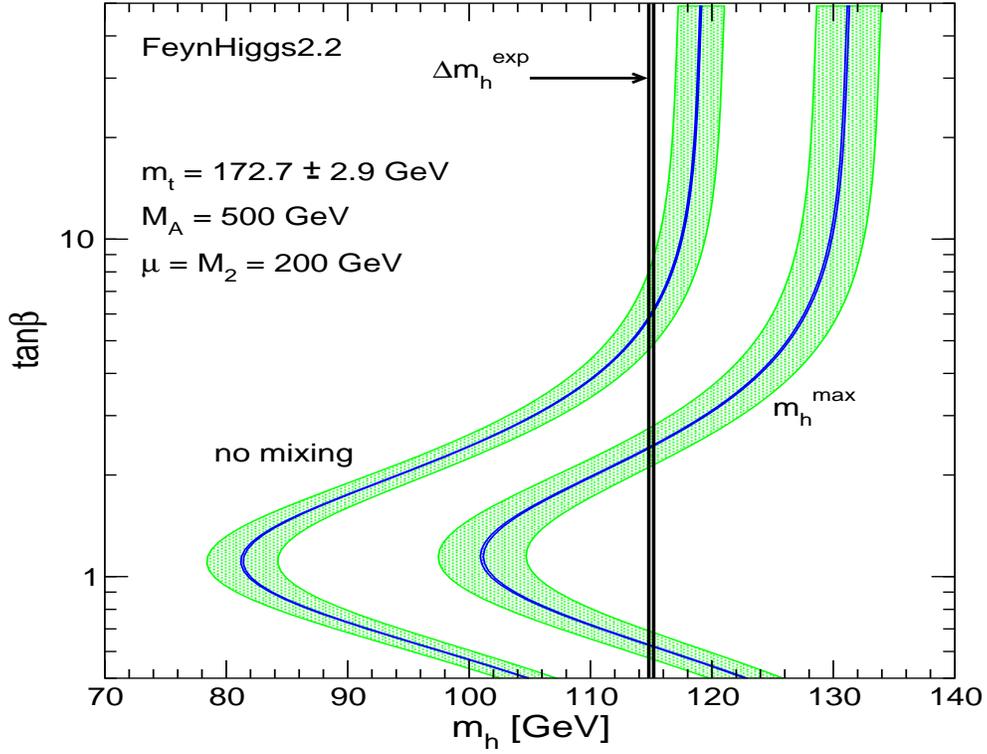}
\caption{
$\mh$ as a function of $\tb$ in the $\mhmax$ and the no-mixing
scenario. The light (green) shaded area corresponds to the current top
quark mass uncertainty of $\de\mtexp = 2.9 \gev$, the dark (blue) shaded
one to the anticipated ILC accuracy, $\de\mtexp = 0.1 \gev$. Also
shown is a possible future $\mh$ measurement at the LHC
of $\de\mh = 0.2 \gev$.
}
\label{fig:mhtb}
\end{center}
\end{figure}

In \reffi{fig:mhtb} 
the effects of the current top quark mass uncertainty on the $\mh$
prediction~\cite{tbexcl} 
are compared to the ILC precision in two benchmark
scenarios, the $\mhmax$ and the no-mixing scenario~\cite{benchmark2}. 
The plot shows $\mh$ as a function of $\tb$, the ratio of the two
vacuum expectation values of the two MSSM Higgs doublets.
Also indicated is a hypothetical $\mh$ measurement at the LHC, while
no intrinsic theoretical uncertainty from unknown higher-order 
corrections is included. Currently this error is estimated to 
$\de\mh^{\rm intr,today} 
\approx 3 \gev$~\cite{PomssmRep,mhiggsAEC,mhiggsWN,habilSH}. In the
future one can hope for an improvement down to 
$\lsim 0.5 \gev$~\cite{PomssmRep,habilSH}. 
If the intrinsic error could be reduced even to $\sim 0.1 \gev$, its
effect in the plot would be roughly as big as the one induced by
$\de\mt = 0.1 \gev$. The inclusion of an intrinsic uncertainty of 
$\sim 1 \gev$ would lead to a significant widening of the 
inner band ($\de\mt$ from ILC) of 
predicted $\mh$ values. In this case the intrinsic uncertainty would
dominate, implying that a reduction of $\de\mt = 1 \gev$ to 
$\de\mt = 0.1 \gev$ would lead only to a moderate improvement of the
overall uncertainty on $\mh$. 

Confronting the theoretical prediction of $\mh$ with a precise
measurement of the Higgs boson mass constitutes a very sensitive test
of the MSSM, which allows to obtain constraints on the model
parameters, in this case $\tb$. However, the sensitivity of the $\mh$
prediction on $\tb$ shown in \reffi{fig:mhtb} cannot directly be
translated into a prospective indirect determination of $\tb$, since
fixed values are assumed for all other SUSY parameters. In a realistic
situation the anticipated experimental errors of the other SUSY
parameters have to be taken into account.
For examples including these parametric errors see \citeres{deltamt,lhcilc}.

%%%%%%%%%%%%%%%%%%%%%%%% F I G U R E %%%%%%%%%%%%%%%%%%%%%%%%%%%%%%%%%%%%%%%%%
%%%%%%%%%%%%%%%%%%%%%%%% F I G U R E %%%%%%%%%%%%%%%%%%%%%%%%%%%%%%%%%%%%%%%%%

%%%%%%%%%%%%%%%%%%%%%%%%%%%%%%%%%%%%%%%%%%%%%%%%%%%%%%%%%%%%%%%%%%%%%%%%%%%%%%%
%%%%%%%%%%%%%%%%%%%%%%%%%%%%%%%%%%%%%%%%%%%%%%%%%%%%%%%%%%%%%%%%%%%%%%%%%%%%%%%

\section{COSMOLOGY}

In this section we focus on the framework of the constrained MSSM
(CMSSM), in which the soft 
supersymmetry-breaking scalar and gaugino masses are each assumed to be
equal at some Grand Unification Theory (GUT) input scale. In this
case, the new independent MSSM 
parameters are just four in number: the universal gaugino mass $m_{1/2}$,
the scalar mass $m_0$, the trilinear soft supersymmetry-breaking parameter
$A_0$, and $\tb$. The 
pseudoscalar Higgs mass $\MA$ and the magnitude of the Higgs mixing 
parameter $\mu$ can be determined by using the electroweak vacuum 
conditions, leaving the sign of $\mu$ as a residual ambiguity.

%%%%%%%%%%%%%%%%%%%%%%%% F I G U R E %%%%%%%%%%%%%%%%%%%%%%%%%%%%%%%%%%%%%%%%%
\begin{figure}[htb!]
\vspace{2em}
\begin{center}
\epsfig{figure=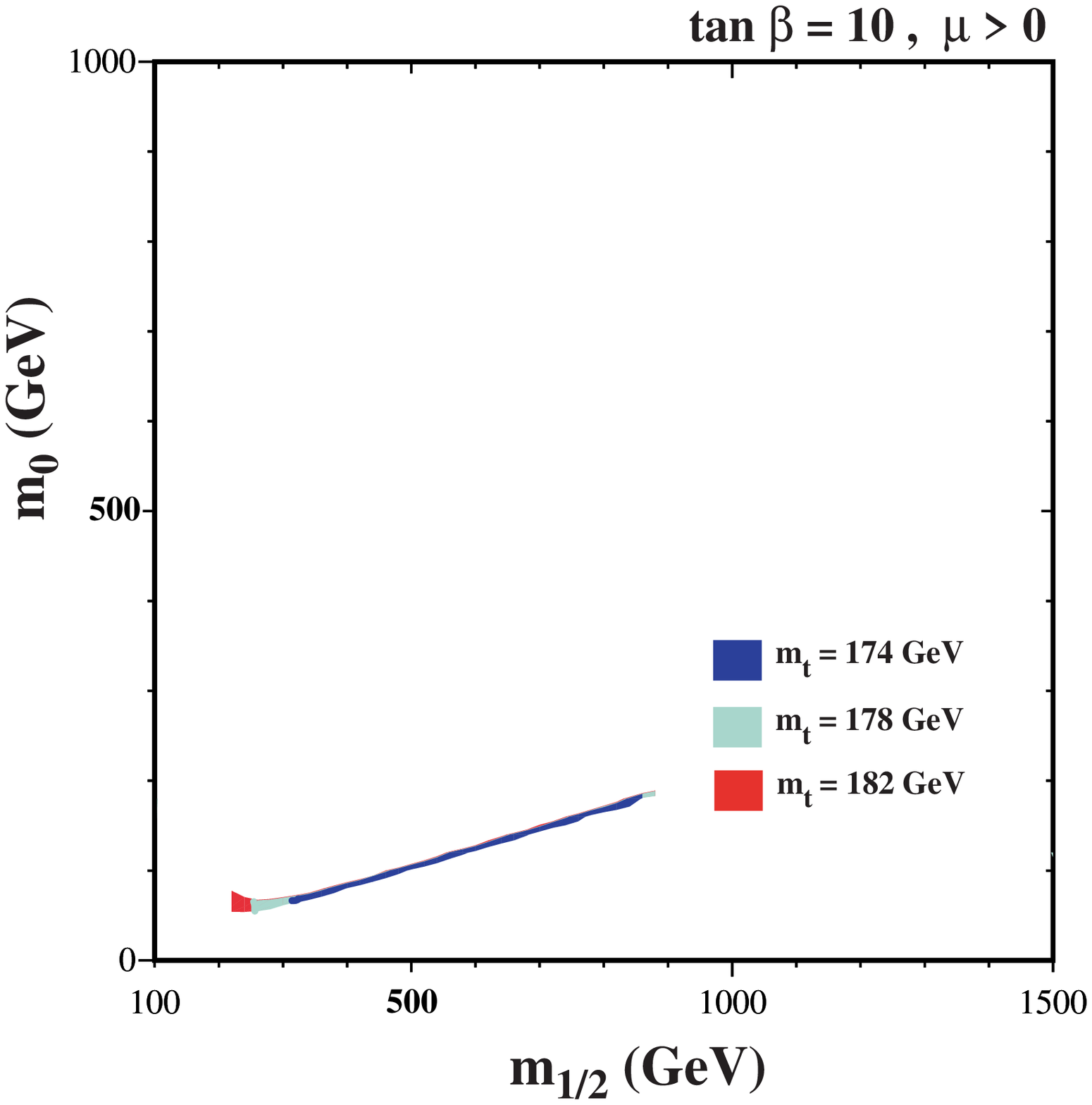, width=7.5cm,height=7.6cm}\qquad
\epsfig{figure=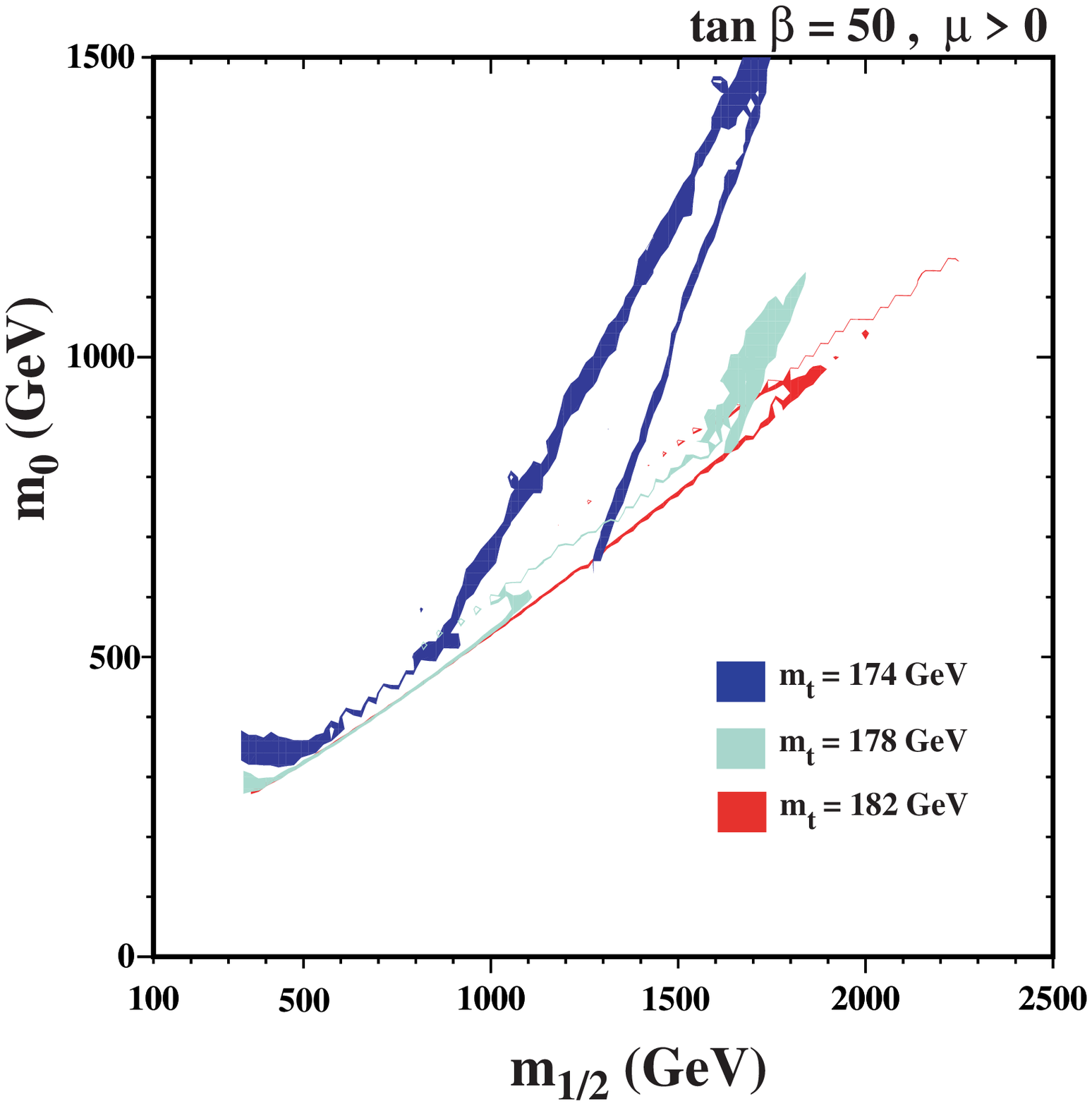, width=7.5cm,height=7.6cm}
\caption{The WMAP strips for $\mu > 0$, $A_0 = 0$
and $\tb = 10$ (left) or $\tb = 50$ (right). The strips are shown for
three top quark mass values, $\mt = 174, 178, 182 \gev$~\cite{ehow3}.}
\label{fig:cdmstrips}
\end{center} 
\end{figure}
%%%%%%%%%%%%%%%%%%%%%%%% F I G U R E %%%%%%%%%%%%%%%%%%%%%%%%%%%%%%%%%%%%%%%%%

The non-discovery of supersymmetric particles and the Higgs boson at 
LEP
and other present-day colliders imposes significant lower bounds on
$m_{1/2}$ and $m_0$. An important further constraint is provided by the
density of cold dark matter (CDM) in the Universe, which is tightly
constrained by 
WMAP and other astrophysical and cosmological data~\cite{WMAP},
leading to $0.094 < \Omega_{\rm CDM} \, h^2 < 0.129$.
This has the effect within the CMSSM, assuming that the dark matter 
consists largely of neutralinos~\cite{EHNOS},
of restricting $m_0$ to very narrow allowed strips
for any specific choice of $A_0$, $\tb$ and the sign of 
$\mu$~\cite{WMAPstrips,wmapothers}.
It is then possible to restrict phenomenological analysis to these
`WMAP strips', see e.g.~\citere{ehow3} and references therein.

Varying the value of $\mt$ has a significant effect on the
regions of CMSSM parameter space allowed by CDM, particularly in the 
`funnels' where neutralinos
annihilate rapidly via the $H, A$ poles. 
Because of the constraints from the anomalous magnetic moment of the
muon~\cite{g-2exp,g-2theo} we focus here on the case with $\mu > 0$.

Plotted in \reffi{fig:cdmstrips}~\cite{ehow3} is the region in the
\plane{m_{1/2}}{m_0} 
for fixed $\tb, A_0$ and $\mu > 0$ for which the relic density
is in the WMAP range. We have applied cuts based on the lower
limit to the Higgs mass~\cite{LEPHiggsSM,LEPHiggsMSSM}, 
$b \to s \gamma$~\cite{pdg}, and require that the LSP be 
a neutralino rather than the stau, see \citere{ehow3} for further
details. The thin strips correspond to the relic density
being determined by either the coannihilation between nearly degenerate
$\tilde \tau$'s and $\chi$'s or, as seen at high $\tb$, by
rapid annihilation when $m_{\chi} \approx \MA/2$.
One can  see in the left plot of \reffi{fig:cdmstrips} that the change
in the WMAP strips for $\mu > 0$ and  $\tb = 10$ is moderate as $\mt$
is varied, reflecting the fact that the allowed strip is dominated by
annihilation of the neutralino  LSP $\chi$ with the lighter stau.
The main effect of
varying $\mt$ is that the truncation at low $m_{1/2}$, due to the Higgs
mass constraint, becomes more important at low $\mt$. This effect is not
visible in the right plot of \reffi{fig:cdmstrips} with $\tb = 50$, where the
cutoff at low $m_{1/2}$ is due to the $b \to s \gamma$ constraint, and
rapid $\chi \chi \to A, H$ annihilation is important at large
$m_{1/2}$. The allowed 
regions at larger $m_{1/2}$ vary significantly with $\mt$ for 
$\tb = 50$, because the $A, H$ masses and hence the rapid-annihilation 
regions are very sensitive to $\mt$ through the renormalization group (RG)
running. 

Thus for large $\tb$ a precise determination of $\mt$ is indispensable
to connect the GUT scale parameters with the CDM measurement. An
uncertainty of $\de\mtexp = 1$--$2 \gev$ would sweep out a large part
of the \plane{m_{1/2}}{m_0}. A precise determination of 
$\de\mtexp = 0.1 \gev$, on the other hand, would result in very thin
and precisely determined strips that give $m_0$ as a function of
$m_{1/2}$ (depending on the precision of $A_0$ and $\tb$ and the
theory uncertainty in the RG running), see also \citere{fawziCDM}.

%%%%%%%%%%%%%%%%%%%%%%%%%%%%%%%%%%%%%%%%%%%%%%%%%%%%%%%%%%%%%%%%%%%%%%%%%%%%%%%
%%%%%%%%%%%%%%%%%%%%%%%%%%%%%%%%%%%%%%%%%%%%%%%%%%%%%%%%%%%%%%%%%%%%%%%%%%%%%%%

\section{CONCLUSIONS}

We have investigated the impact of the experimental error of the top
quark mass on various physics scenarios. Especially we have compared
the parametric error induced by the LHC uncertainty of 
$\de\mt^{\rm exp,LHC} \approx \mbox{1--2} \gev$ 
with the one of the ILC, $\de\mt^{\rm exp,ILC} \approx 0.1 \gev$. 

Concerning electroweak precision observables such as $\MW$ and
$\sweff$ the parametric error induced by $\de\mtexp$ has been
investigated. It will match the intrinsic error of $\MW$ and
$\sweff$ and their other parametric errors only if the ILC precision
of $\de\mtexp \sim 0.1 \gev$ can be 
reached. Otherwise the parametric error from $\mt$ will dominate the future
uncertainties and hamper the otherwise powerful consistency checks of
the model under investigation. 

The large Yukawa coupling of the top quark can induce large
corrections to the prediction of the Higgs boson mass and result in
corresponding parametric uncertainties. 
As has been discussed for the specific example of the MSSM, 
the prospective experimental error of $\mh$
at the LHC can only be matched if $\de\mtexp \sim 0.1 \gev$ can be achieved. 

Furthermore, the prediction of the cold dark matter abundance in the CMSSM
has been analyzed. An $\mt$ uncertainty enters the CDM prediction
in particular through renormalization group running from the GUT scale to the
electroweak scale. For large values of $\tb$ the $\mt$ error can
induce a large uncertainty in the CDM prediction. The precision of 
$\de\mtexp \sim 0.1 \gev$ is desirable for fully exploiting the
restrictions of the CDM measurements on the CMSSM
parameter space, permitting thorough consistency checks of the model.

Summarizing, the already impressive LHC precision on $\mt$ will not be
sufficient to match the required future precisions in various physics
models and scenarios. Only the ILC precision will be able to reach the
desired accuracy.

%%%%%%%%%%%%%%%%%%%%%%%%%%%%%%%%%%%%%%%%%%%%%%%%%%%%%%%%%%%%%%%%%%%%%%%%%%%%%%%
%%%%%%%%%%%%%%%%%%%%%%%%%%%%%%%%%%%%%%%%%%%%%%%%%%%%%%%%%%%%%%%%%%%%%%%%%%%%%%%

\subsection*{Acknowledgments} 

We thank J.~Ellis, J.~Haestier, S.~Kraml, K.~Olive, W.~Porod and 
D.~St\"ockinger for collaboration on various aspects
presented here.

%%%%%%%%%%%%%%%%%%%%%%%%%%%%%%%%%%%%%%%%%%%%%%%%%%%%%%%%%%%%%%%%%%%%%%%%%%%%%%%
%%%%%%%%%%%%%%%%%%%%%%%%%%%%%%%%%%%%%%%%%%%%%%%%%%%%%%%%%%%%%%%%%%%%%%%%%%%%%%%

%%%%%%%%%%%%%%%%%%%%%%%%%%%%%%%%%%%%%%%%%%%%%%%%%%%%%%%%%%%%%%%%%%%%%%%%%%%%%%%
%%%%%%%%%%%%%%%%%%%%%%%%%%%%%%%%%%%%%%%%%%%%%%%%%%%%%%%%%%%%%%%%%%%%%%%%%%%%%%%

\end{document}